\documentclass[%
 aip,
 jmp,%
 amsmath,amssymb,
reprint,%
]{revtex4-1}

\usepackage{graphicx}
\usepackage{dcolumn}
\usepackage{bm}

\newcommand{\R}{{\mathbb R}}

\begin{document}

\preprint{AIP/123-QED}

\title[Subexponential instability implies ...]{Subexponential instability implies infinite invariant measure}

\author{Takuma Akimoto}
\email{akimoto@aoni.waseda.jp}
\author{Yoji Aizawa}%
\affiliation{ Department of Applied Physics, Advanced School of Science and Engineering, Waseda University, Okubo 3-4-1, Shinjuku-ku, Tokyo 169-8555, Japan.
}%

\date{\today}

\begin{abstract}
We study  subexponential instability to characterize a dynamical instability of weak chaos. We show that a dynamical system with  subexponential instability has an infinite invariant measure, and then we present the generalized Lyapunov exponent to characterize  subexponential instability. 
\end{abstract}

\pacs{05.45.Ac, 05.20.-y, 05.40.-a}
\keywords{Weak chaos, subexponential instability, infinite invariant measure}
\maketitle

\begin{quotation}
Chaotic dynamical systems with the zero Lyapunov exponent have attracted much interest in non-equilibrium statistical mechanics. Chaos in such dynamical systems is called weak chaos. We consider the Pomeau-Manneville map as an example of weak chaos, whose dynamical instability is not exponential one. Then, we define subexponential instability using the average of the logarithm of the separation of two nearby orbits. A dynamical instability of weak chaos is characterized as subexponential instability. We demonstrate the relation between subexponential instability and infinite invariant measure. In particular, we show that subexponential instability implies infinite invariant measure. The generalized Lyapunov exponent is studied to quantify a dynamical instability of weak chaos.
\end{quotation}

\section{Introduction}

In the foundation of statistical mechanics,  chaos plays an important role in deriving a stochastic description from  microscopic dynamics \cite{Dorfmann1999}. The existence of an equilibrium state in  dynamical systems requires  ergodicity, i.e., time average being equal to space average. This is  because macroscopic quantities, which result from time averages of  microscopic observation functions, are almost equal to space average (constant). 
\par
Non-equilibrium phenomena can be classified into two different states: non-equilibrium stationary state and non-equilibrium non-stationary state. In non-equilibrium stationary state, the time average of an observation function is almost constant but it depends on subsystems. In other words, macroscopic observables restricted to subsystems are almost constant but depend on the position of subsystems. On the other hand, macroscopic observables are not constant in non-equilibrium non-stationary state. For example, an approach to an equilibrium state such as a diffusion is non-equilibrium non-stationary state because macroscopic observables depend on time. Other examples are  random processes such as turbulence and chemical reaction, where macroscopic observables do not approach  an equilibrium state but change randomly in time. In such phenomena, the time average of an observation function in underlying microscopic dynamics need to be intrinsically random. However,
the foundation of non-equilibrium statistical mechanics on the basis of time average has not been studied at all. \par
 Recently, random behaviors of physical quantities in non-equilibrium non-stationary phenomena such as anomalous diffusions and intermittency have been studied extensively in both experiments and theories \cite{Golding2006, Brok2003, He2008, Akimoto2008}. In particular, randomness of the generalized diffusion coefficient has been clearly observed in the diffusion of mRNA molecules inside the cell \cite{Golding2006}. Since then, this result has been elucidated theoretically through the continuous time random walk model of anomalous diffusion \cite{He2008}. In general, anomalous diffusion processes can be represented by deterministic dynamical systems \cite{Geisel1984}. It has been known that dynamical systems describing subdiffusions, where mean square displacement increases as $\langle x(t)^2 \rangle = o(t)$, show aging and infinite invariant measure \cite{Barkai2003}.
Infinite measure systems, i.e., dynamical systems with infinite invariant measures, have  attracted much interest to found the non-equilibrium statistical mechanics \cite{
 Aizawa1989a,Akimoto2008,Akimoto2008a,Korabel2009}. The notable point in infinite measure systems is that the time average of an observation function is intrinsically random \cite{Aaronson1981, Akimoto2008}. More precisely, the time average of an observation function converges in distribution, and its distribution depends on the class of the observation function. This is reminiscent of the randomness of the time average in non-equilibrium non-stationary states. \par
 In this paper, we develop the concept of ergodicity toward the foundation of non-equilibrium statistical mechanics based on  time averages in
dynamical systems. Our aim is to characterize the dynamical instability of infinite measure systems.   In finite measure systems a dynamical instability can be characterized by the Lyapunov exponent. In a one-dimensional map $T(x)$, the Lyapunov exponent defined by the time average of $\ln |T'(x)|$ indicates the exponential rate of the separation between two nearby orbits. By Birkhoff's individual ergodic theorem, the time average of $\ln |T'(x)|$ converges to the average of $\ln |T'(x)|$ with respect to the invariant measure for almost all initial points  \cite{Birkhoff1931}.  Here, we focus on the subexponential instability which is characterized by the subexponential growth of the separation of two nearby orbits. Accordingly, the Lyapunov exponent converges to zero in dynamical systems with  subexponential instability. We show that dynamical systems with subexponential instability have  infinite invariant measures. We then study the generalized Lyapunov exponent defined as the average growth rate of  subexponential separation.


\section{Subexponential instability}

Gaspard and Wang suggested that the separation between nearby trajectories  shows stretched exponential growth in Pomeau-Mannivelle maps with  infinite invariant measures \cite{Gaspard1988}. However, the dependence of the growth rate on an initial point was not studied.\par
 Based on the Darling and Kac theorem \cite{Darling1957}, Aaronson proved that the time average of the positive $L^1(m)$ function \cite{FN1}  
converges in distribution, where $m$ is an invariant measure. That is, the growth rate strongly depends on an initial point and is intrinsically random. When the logarithm of the derivative of a map is the positive $L^1(m)$ function, the normalized Lyapunov exponent converges in distribution:
\begin{equation}
\Pr\left\{
\frac{1}{a_n}\sum_{k=0}^{n-1} \ln |T'(x_k)| \leq t\right\}
\rightarrow M_{\alpha}(t) \quad {\rm as}~n\rightarrow \infty,
\label{eq:1}
\end{equation}
where $a_n$ is called the return sequence determined by the map and $M_{\alpha}(t)$ is the normalized Mittag-Leffler distribution \cite{FN2}. 
Generally, the return sequence is determined using the {\it wandering rate} defined by
\begin{equation}
w_n=m\left(\bigcup_{k=0}^n T^{-k}A\right),
\end{equation}
where $0<m(A)<\infty$. In particular, the return sequence is given by
\begin{equation}
a_n \sim \frac{n}{\Gamma(1+\alpha)\Gamma(2-\alpha)w_n},
\end{equation}
where $w_n$ is regularly varying at $\infty$ with index $\alpha$ \cite{Aaronson1997}.
 For example, the sequence $a_n$ of the Pomeau-Manneville map,
\begin{equation}
x_{n+1}=
T(x_n)=x_n+cx_n^{1+p}\quad ({\rm mod}~1),
\end{equation}
where $c>0, p\geq 1$, is given by
\begin{equation}
a_n\propto\left\{
\begin{array}{ll}
n^{\frac{1}{p}} \quad &(p>1)\\
\\
{\displaystyle \frac{n}{\log n} }&(p=1).
\end{array}
\right.
\end{equation}
Because of $\ln |T'(x)|\in L^1(m)$, the distribution of the normalized Lyapunov exponent  converges to a Mittag-Leffler distribution of order $\alpha$, where the order $\alpha$ is given by $1/p$. Figure 1 shows that the distribution of the normalized Lyapunov exponent, 
$
\lambda=\frac{1}{a_n}\sum_{k=0}^{n-1} \ln |T'(x_k)|,
$
 converges to a Mittag-Leffler distribution even when the time is finite $(n=10^8)$.
\par

\begin{figure}
\includegraphics[height=.9\linewidth, angle=-90]{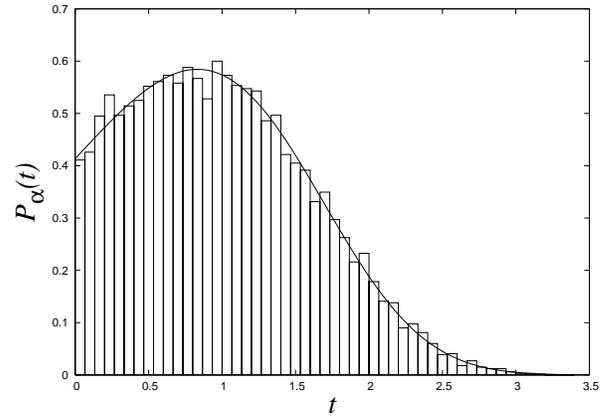}
\caption{Probability density function $P_{\alpha}(t)$ of the normalized Lyapunov exponent of the Pomeau-Mannivelle map $(p=1.5)$. The sequence $a_n$ is given by $a_n=\langle \lambda \rangle n^{\alpha}$, where $\alpha=2/3$ and $\langle \lambda \rangle=2.622$.}
\end{figure}
We consider the separation of two nearby orbits ($x_n$ and $x_n'$) defined as
\begin{equation}
\Delta_n\equiv\lim_{\Delta x(0)\rightarrow 0}\left|\frac{\Delta x(n)}{\Delta x(0)}\right|
=\left|\mathop{\prod}_{k=0}^{n-1}T'(x_k)\right|,
\end{equation}
where $\Delta x(n)=x_n-x'_n$. 
 By (\ref{eq:1}), the distribution for 
the growth of the separation can be given by
\begin{equation}
\Pr \left\{ \Delta_n\leq t\right\}
 \rightarrow M_{\alpha}\left(\frac{\ln t}{a_n}\right)\quad {\rm as}~n\rightarrow \infty.
\end{equation}
For large $n$, the average of the separation is given as
\begin{equation}
\left\langle \Delta_n
\right\rangle \cong 
\langle e^{a_nY_{\alpha}} \rangle_{ML},
\end{equation}
where 
  $\langle \cdot \rangle$ represents the average with respect to an initial ensemble, 
$\langle \cdot \rangle_{ML}$ represents the average with respect to a 
Mittag-Leffler distribution, and $Y_{\alpha}$ is a random variable with a Mittag-Leffler distribution of order $\alpha$.
Accordingly, one can write 
\begin{equation}
\langle e^{a_n Y} \rangle_{ML}
=\sum_{k=0}^{\infty} \frac{\Gamma(1+\alpha)^k}{\Gamma(1+k\alpha)}a_n^k.
\end{equation}
We then have
\begin{eqnarray}
\langle e^{a_n Y} \rangle_{ML} &>&\sum_{k=0}^{\infty}
\frac{(\gamma a_n)^k}{(\lfloor k\alpha\rfloor+1)!}\nonumber\\
&>& \frac{(\gamma a_n)^{\lfloor p \rfloor+1}-1}{\gamma a_n-1}\exp[(\gamma a_n)^{\lfloor p \rfloor}],
\end{eqnarray}
where $\gamma =\Gamma (1+\alpha)$. The upper bound is obtained as
\begin{equation}
\frac{1}{n}\ln \Delta_n =\frac{1}{n}\sum_{k=0}^{n-1} \ln |T'(x_k)| \leq M,
\end{equation}
where $M=\ln [1+c(1+p)]$.
 For large $n$, we have the following inequality:
\begin{equation}
 \exp(B n^{\lfloor p \rfloor\alpha})<\left\langle \Delta_n \right\rangle <
  \exp(M n),
\end{equation}
where $B$ is constant.
As shown in Fig. 2, the fluctuation of the separation $\Delta_n$ is extremely large and the average of $\Delta_n$ is  almost determined by the largest value of $\ln \Delta_n$. This implies that  subexponential instability should be characterized as the average of the logarithm of the separation, $\langle \ln \Delta_n \rangle$,
rather than that of the separation, $\langle \Delta_n \rangle$. 
We call that {\it a dynamical system has subexponential instability} if there exists the sequence $a_n=o(n)$  such that
\begin{equation}
\left\langle \frac{1}{a_n}\sum_{k=0}^{n-1} \ln |T'(x_k)| \right\rangle\rightarrow 1 \quad 
{\rm as}~n\rightarrow \infty.
\end{equation}

\begin{figure}
\includegraphics[height=.9\linewidth, angle=-90]{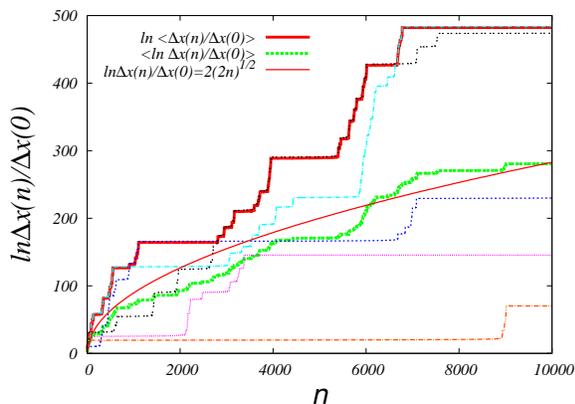}
\caption{(color online). Numerical simulations for the growth of the separation in Boole transformation (23). The  separations $|T'(x_n)\cdot T'(x_{n-1})\cdots T'(x_1)|$ for five different initial points are represented by five dotted lines, the logarithm of the average of $\Delta_n$ and the average of the logarithm of $\Delta_n$ are represented by the thick line and the thick dotted line, and the theoretical curve Eq. (26) is also represented by the line.}
\end{figure}

\section{Subexponential instability implies infinite invariant measure}

We show that invariant measure can not be normalized   when a dynamical system has subexponential instability, and then that the normalized Lyapunov exponent becomes intrinsically random. Here, we consider a one-dimensional conservative, ergodic, measure preserving map with an invariant measure absolutely continuous with respect to Lebesgue measure in the sense of infinite ergodic theory\cite{Aaronson1997}.
In other words, we do not consider dynamical systems with the trivial zero Lyapunov exponents such as $T'(x)=1$ for almost all $x$. Moreover, we do not consider dynamical systems where there exist attractive indifferent periodic points, $(T^{m}(x_k^*))'=1$ for period $m$ points $x_1^*, \cdots, x^*_{m}$ $(k=1,\cdots , m)$ in the sense of Milnor\cite{Milnor1985}. The typical example of such dynamical systems is the dynamical system {\it at the period doubling bifurcation points}. At the critical point $(m\rightarrow \infty)$, $\langle \Delta_n\rangle$ increases as power law \cite{Grassberger1981}. However, we do not consider this case because there is no sensitive dependence on initial conditions nor an invariant measure absolutely continuous with respect to Lebesgue measure \cite{Collet1980}. We note that there is at most one $\sigma$-finite invariant measure in a one-dimensional conservative, ergodic, measure preserving map.\par

First, we assume $X_k=\ln T'(x_k)\geq 0$. If $X_1,X_2,\cdots$ are independent and identically distributed ($i.i.d.$) random variables, then by the laws of large numbers we have
\begin{equation}
\Pr \left\{ \left|\frac{X_1+\cdots X_n}{n}- E(X_k) \right|
>\epsilon \right\} \rightarrow  0 \quad 
{\rm as}~n\rightarrow \infty,
\end{equation}
where $E(X)(>0)$ is the expectation of $X$, which contradicts  subexponential instability  ($E(X_1+\cdots +X_{n})=o(n)$). This implies that $X_k$ is not an $i.i.d.$ random variable. Moreover, by $E(X_1+\cdots +X_{n})=o(n)$ 
we have
\begin{equation}
E(X_n)\rightarrow 0\quad {\rm as}~n\rightarrow \infty.
\end{equation}
It follows that 
the probability $\Pr \{X_k\leq \epsilon\}$ increases with time for all $\epsilon>0$. That is, $x_k$ is not an $i.i.d.$ random variable, but  trapped  near the indifferent fixed point ($T'(a)=1$) for a very long time \cite{FN3}, where $a$ $(<\infty)$ is an indifferent fixed point. 
 In other words, the empirical distribution of an orbit $\{x_k\}$ converges to the delta distribution: 
For all $d>b>a$, 
\begin{equation}
\int_b^d \rho_n(x)dx \rightarrow 0 \quad 
{\rm as}~n\rightarrow \infty,
\end{equation}
where
\begin{equation}
\rho_n(x) =
\frac{\delta_{x_0}(x)+\delta_{x_1}(x)+\cdots +\delta_{x_n}(x)}{n}.
\end{equation}
Therefore, the invariant measure is an infinite one. This is because there exists a unique invariant measure absolutely continuous with respect to Lebesgue measure and the empirical distribution of an orbit converges to the invariant measure if the map has a finite invariant measure ({\it probability measure}).\par 
 As will be seen later, an indifferent fixed point of Boole transformation is at $\pm\infty$. In the case of $T'(\pm\infty)=1$, a dynamical system is an open system, i.e, the space of a dynamical system is not compact. By the condition of subexponential instability, $E(X_1+\cdots +X_{n})=o(n)$, 
Eq. (16) is fulfilled for $-\infty <b<d<\infty$. Therefore, the invariant measure is an infinite one. This impossibility of the normalization 
of the invariant measure stems from the non-compactness of the space of a dynamical system. Performing a change of a  variable, we can obtain a compact dynamical system with  indifferent fixed points \cite{Thaler1983} (see Appendix A). Intuitively speaking, the infinite invariant measure on the non-compact support is obtained by stretching the infinite invariant measure  on the compact support near the indifferent fixed points.
\par
Next, we consider the case of $X_k\in \R$, $E(|X_1|+\cdots +|X_n|) =O(n)$ and $E(X_1+\cdots +X_n)=o(n)$ \cite{FN4}. 
 Let $X_1,X_2,\cdots$ be $i.i.d.$ random variables with the mean 0. Then, there exists the sequence $a_n$ such that the limit distribution of 
 $(X_1+\cdots +X_n)/a_n$ converges to be a normal or a stable distribution \cite{Feller1971}. It follows that the probability of the negative Lyapunov exponent ($\lambda<0$) remains positive and that the Lyapunov exponent converges to zero as $n$ goes to infinity. This means that the dynamical system has an attractive indifferent periodic points. As we mentioned above, we do not consider this dynamical system because there is no sensitivity to initial conditions. Therefore,  $X_1,X_2,\cdots$ are not $i.i.d.$ random variables with the mean 0.\par 
 There are two cases: (i) The expectation of $X_n$ is positive and $E(X_n)\rightarrow 0$ as $n\rightarrow \infty$. In a similar way
we can show that the invariant measure is infinite.  
 (ii) Due to canceling out of $X_1+\cdots + X_n$, $E(X_1+\cdots +X_n)$ does not grow linearly. We introduce some notations (see Fig. 3): the partial sum of $X_n$, $S_n=X_1+\cdots +X_n$, the maximum of $S_n$, $m_n=\max \{0, S_1, \cdots, S_n\}$, the number of renewals of maximum, $N_n=\min\{k : m_k>m_{k-1}, m_k>m_{k+1}, k>N_{n-1}\}$, and the $n$th maximum, $M_n=m_{N_n}$.
 \begin{figure}
\includegraphics[height=.4\linewidth, angle=0]{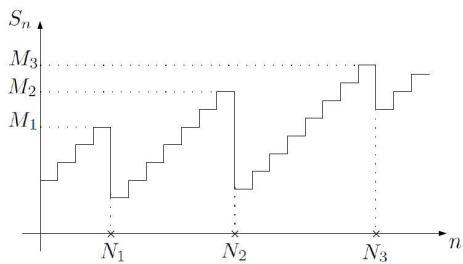}
\caption{ Schematic illustration of $N_n$ and $M_n$. }
\end{figure}
 The condition $E(X_1+\cdots +X_n)=o(n)$ leads to $E(N_n-N_{n-1})=\infty$  or $E(M_n-M_{n-1})\rightarrow 0$ as $n\rightarrow \infty$, where the interevent times, $N_n-N_{n-1}$, are {\it i.i.d.} random variables.\par 
  In the case of $E(N_n-N_{n-1})=\infty$ and $E(M_n-M_{n-1})<\infty$,  an orbit is occasionally trapped in the region  near the fixed point for a very long time because the residence time in the region near the fixed point becomes larger as the reinjection is closer to the fixed point. 
 Since the probability density function of the residence time on $[a-\epsilon,a+\epsilon]$ for the tail part is invariant for all $\epsilon$, the expectation of the residence time on $[a-\epsilon,a+\epsilon]$ denoted by $T_{A_{\epsilon}}$ is infinite, where $a$ is the fixed point. By the law of large numbers,  we have
 \begin{equation}
\frac{1}{n}\sum_{k=0}^{n-1} 1_{[a-\epsilon,a+\epsilon]}(x_k) \rightarrow \frac{T_{A_{\epsilon}}}{T_{A_{\epsilon}}+T_{A_{\epsilon}^c}}=1,
\end{equation}
for all $\epsilon$ as $n\rightarrow \infty$, where  $T_{A_{\epsilon}^c}$ is the expectation of the residence time on the complement of the set $[a-\epsilon,a+\epsilon]$. This means that the invariant measure is an infinite one. 
It is noteworthy that the fixed point does not need to be the indifferent fixed point (see Fig. 4). As shown in Appendix B, the transformation with a flat critical point such as
\begin{equation}
T(x)=\left\{
\begin{array}{ll}
 2x &
 x\in [0,1/2]\\
\noalign{\vskip0.2cm}
\exp (-\frac{1}{x-1/2}+2)/2 &
 x\in (1/2,1].
\end{array} \right.
\end{equation}
has  subexponential instability with $\ln |T'(x)| \in \R$ and an infinite invariant measure \cite{Inoue1992, Zweimuller2004}, which is compatible with the above result.\par 
\begin{figure}
\includegraphics[height=.9\linewidth, angle=-90]{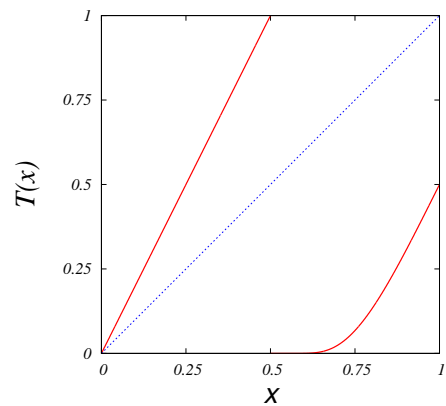}
\caption{(Color online). The transformation with a flat critical point. The flat critical point is at $x=0.5+0$.}
\end{figure}
  In the  case of  $E(M_n-M_{n-1})\rightarrow 0$ as 
$n\rightarrow \infty$ and $E(N_n-N_{n-1})<\infty$, this leads to
 $(T^{(N_n-N_{n-1})}(x_{N_{n-1}}))'\rightarrow 1$ as $n\rightarrow \infty$. It follows that an orbit is trapped near the indifferent periodic points. Hence, the invariant measure is an infinite one. Modifying the logistic map at the  tangent bifurcation point to change the one-sided stable indifferent periodic points   to the repulsive indifferent periodic points, we have a dynamical system with an infinite invariant measure. In particular, the modified logistic map is given by
 \begin{equation}
T(x)=\left\{
\begin{array}{ll}
 -2(x-e)+(1+2\sqrt{2})e(1-e) &
 x\in [0,e]\\
\noalign{\vskip0.2cm}
(1+2\sqrt{2})x(1-x) &
 x\in (e,1],
\end{array} \right.
\end{equation}
where $T^3(e)=e$ $(e<0.2)$. The three iterated map $T^3(x)$ has three indifferent fixed points (Fig. 5). In particular,  $T^3(x)=x+C(x-e)^2+o((x-e)^2)$ as $x\rightarrow e+0$, where $C\geq 0$ is constant. Therefore, the invariant measure is infinite one \cite{Thaler1983}.
\par

When $\ln |T'(x)|$ is the positive $L^1(m)$ function, by the Darling-Kac and Aaronson theorem, the normalized Lyapunov exponent converges in distribution to a Mittag-Leffler distribution.
On the other hand, when $\ln |T'(x)|$ is $L_{loc}^1(m)$ with finite mean function, the Lyapunov exponent converges in distribution to the generalized arcsine distribution \cite{Akimoto2008}. When $\ln |T'(x)|$ is $L_{loc}^1(m)$ with infinite mean function, the Lyapunov exponent converges in distribution to a stable distribution \cite{Akimoto2008}. However, these cases do not show  subexponential instability. Therefore, the
subexponential instability makes the Lyapunov exponent intrinsically random, and the normalized Lyapunov exponent could possibly be described using a Mittag-Leffler distribution.

\begin{figure}
\includegraphics[height=.9\linewidth, angle=-90]{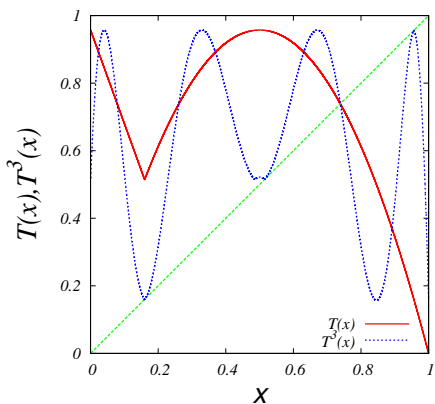}
\caption{(Color online). The modified logistic map. The solid line and the dotted line represent $T(x)$ and $T^3(x)$, respectively.}
\end{figure}

\section{Generalized Lyapunov exponent}
Indicators characterizing weak chaos have been studied by quantifying the complexity or the sensitivity to initial conditions \cite{Korabel2009, Galatolo2003, Bonanno2004}. To characterize  subexponential instability we use
 the generalized Lyapunov exponent proposed by Korabel and Barkai \cite{Korabel2009}. The generalized Lyapunov exponent $\lambda(a_n)$ is defined as 
\begin{equation}
\lambda(a_n)=\left\langle \frac{1}{a_n}\sum_{k=0}^{n-1} \ln |T'(x_k)| \right\rangle,
\end{equation}
where $\langle \cdot \rangle$ represents the average with respect to  initial points. It is remarkable that one can choose an arbitrary initial ensemble if it is continuous with respect to  Lebesgue measure.  Note that there exists the sequence $a_n$ such that $0<\lambda (a_n)<\infty$ as $n\rightarrow \infty$ in dynamical systems with subexponential instability.
The sequence $a_n$ presents a global instability of a dynamical instability and ${\displaystyle \lambda=\lim_{n\rightarrow\infty} \lambda (a_n)}$ represents the rate of a subexponential growth. We call $(a_n,\lambda)$ the {\it Lyapunov pair}.\par
We review the Darling-Kac and Aaronson theorem. Let $T$ be a transformation on $X$ with  subexponential instability, and let $\ln |T'(x)|$ be the positive $L^1(m)$ function. Then the normalized Lyapunov exponent converges in distribution to a Mittag-Leffler distribution: There exists $a_n$ such that
\begin{equation}
\Pr\left\{
\frac{1}{a_n}\sum_{k=0}^{n-1} \ln |T'(x_k)| \leq t\right\}
\rightarrow M_{\alpha}(t) \int_X \ln |T'(x)|dm 
\end{equation}
as $n\rightarrow \infty$.
\par
We demonstrate  subexponential instability for Boole transformation. Boole transformation $T: \R \rightarrow \R$ is given by 
\begin{equation}
x_{n+1}=T(x_n)=x_n-\frac{1}{x_n}.
\end{equation} 
The invariant measure is uniform, i.e., of infinite measure. The return sequence $a_n$ is given as \cite{Aaronson1997}
\begin{equation}
a_n=\frac{\sqrt{2n}}{\pi}.
\end{equation}

By the Darling-Kac and Aaronson theorem, we can obtain the generalized Lyapunov exponent 
\begin{equation}
\lambda(\sqrt{2n}/\pi) = \int_{-\infty}^{\infty} \ln |1+1/x^2|dx=2\pi.
\end{equation}
Thus, the subexponential instability is represented by
\begin{equation}
\left\langle \ln \Delta_n \right\rangle\sim
2\sqrt{2n}\quad{\rm as}~n\rightarrow \infty.
\end{equation}

\section{Conclusions}
In conclusion,
we have shown that dynamical systems with subexponential instability have infinite invariant measures. It follows that a dynamical system with a finite invariant measure does not show subexponential instability. Considering Birkhoff's ergodic theorem, we see that a dynamical system has the positive Lyapunov exponent if the invariant measure $m$ is a probability measure and  the function $\ln|T'(x)|$ is an $L^1(m)$ function. Furthermore, we have presented that one dimensional maps with the {\it non-trivial zero Lyapunov exponent} are described as infinite measure systems such as
transformations with indifferent fixed or periodic points which are repellers and transformations with flat critical points. 
\par  
While it has been known that random dynamical systems have aperiodic orbits with the negative Lyapunov exponent \cite{Wang2004}, the relation between an invariant measure and the Lyapunov exponent is still unknown in random dynamical systems. However, the invariant measure becomes infinite when the Lyapunov exponent is zero in the drift-diffusion model \cite{Ashwin1998}.
Moreover, the transformation $Tx=Dx$ with probability 1/2 and $Tx=x/D$ with probability 1/2 does not have attractors but have an infinite invariant measure absolutely continuous with respect to Lebesgue measure.
We conjecture that the {\it non-trivial zero Lyapunov exponent} would imply an infinite invariant measure. 
\par
Finally, we characterized the separation of nearby orbit as the generalized Lyapunov exponent. There are different definitions of the generalized Lyapunov exponent \cite{Fujisaka1983, Badii1988, Mori1989}. Kuzovkov {\it et al.} applied the generalized Lyapunov exponent to the Anderson localization problem \cite{Kuzovkov2002}.
Compared to these definitions, the generalized Lyapunov exponent in this paper focuses on weak instabilities such as stretched exponential and algebraic instabilities. 
 It would be interesting to make clear the relation between the macroscopic observables in non-equilibrium state and the generalized Lyapunov exponent. We expect that the generalized Lyapunov exponent will be useful to understand the non-equilibrium non-stationary phenomena. In particular, we hope that  random behaviors of transport coefficients in anomalous diffusions will be characterized by the generalized Lyapunov exponent.

\begin{acknowledgments}
The authors wish to thank Tomoki Inoue, Takahito Mitsui, Masaki Nakagawa, Soya Shinkai and Nathan Weissburg for fruitful discussions. 
We would also like to thank Nickolay Korabel and Eli Barkai for reviewing the manuscript and discussing the results.
This research was financially supported by the Waseda University Grant for Special Research Projects.
\end{acknowledgments}

\appendix

\section{Change of a variable}
In Boole transformation a change of a variable, ${\displaystyle \phi (x)=\frac{x-2+\sqrt{4+x^2}}{2x}}$, yields a map with indifferent fixed points on $[0,1]$. The transformed map $S: [0,1]\rightarrow [0,1]$, 
\begin{equation}
S(x)=\phi \circ T\circ \phi^{-1}(x),
\end{equation}
is given by
\begin{equation}
S(x)=\left\{
\begin{array}{ll}
{\displaystyle \frac{x(1-x)}{1-x-x^2}}\quad &x\in [0,1/2]\\
\\
{\displaystyle \frac{1-2x}{1-3x+x^2}} &x\in (1/2,1].
\end{array}
\right.
\end{equation}
This map has two indifferent fixed points at $x=0$ and 1, and 
the asymptotic behavior at $x=0$ is given by 
\begin{equation}
S(x) \cong x+x^3+o(x^3).
\end{equation}
Hence, we see  that the invariant measure is an infinite one.

\section{Subexponential instability of a map (19)}
Due to the flat critical point, the invariant measure is an infinite invariant measure and the invariant density is not bounded at $x=0$. 
The inequality, $2^{-N-1}<T(x)<2^{-N}$, is fulfilled when the residence time in $[0,1/2]$ is $N$ and $x\in (1/2,1]$.
This inequality is solved as
\begin{equation}
\frac{1}{N\ln 2 +2}+\frac{1}{2}<x<\frac{1}{(N-1) \ln 2+2}+\frac{1}{2}.
\end{equation}
Using
\begin{equation}
\ln |T'(x)|=\left\{
\begin{array}{ll}
\ln 2 \quad &x\in [0,1/2]\\
\\
{\displaystyle \frac{-1}{x-\frac{1}{2}}+2-\ln 2 + \ln (x-\frac{1}{2})^{-2}} &x\in (1/2,1],
\end{array}
\right.
\end{equation}
we have the inequality for the partial sum $S_{N+1}=X_1+\cdots +X_{N+1}$:
\begin{eqnarray*}
2 \ln \{(N-1)\ln 2+2\}
 &<& S_{N+1}=\ln |T'(x)| +N\ln 2\\
 &<&-\ln 2+2 \ln (N\ln 2+2).
\end{eqnarray*}
Let $l_n$ denote the number of reinjections on $[0,1/2]$ until $n$ iterations.  Since the distribution of the injection on $[1/2,1]$ is uniform, the distribution of the residence time in the interval $[0,1/2]$
 is given as 
\begin{eqnarray}
F(N)=\Pr\{N_k \leq N\} &=&\Pr \{ T(x) \leq 2^{-N}\} \nonumber\\
&=& \Pr \left\{x \geq \frac{1}{N\ln 2 +2}+\frac{1}{2}\right\}\nonumber\\
&=& 1-\frac{2}{2+N\ln 2}.
\end{eqnarray}
Therefore, 
\begin{equation}
E(N_k)=\int_0^{\infty} NF'(N)dN> \frac{1}{2+\ln 2}\int_1^{\infty}\frac{1}{N}dN=\infty.
\end{equation}
By the renewal theory \cite{Feller1971}, there exists a constant  $K>0$ such that
\begin{equation}
 l_n = K\frac{n}{\ln n}+o\left(\frac{n}{\ln n}\right)\quad {\rm as}~n\rightarrow \infty.
\end{equation} 
  Using the arithmetic and geometric means inequality, we have
\begin{eqnarray}
E(X_1+\cdots +X_n)&<&E(S_{N_1+1}+\cdots + S_{N_{l_n}+1})\nonumber\\
&<& 2 E[\ln \{(N_1+1) \cdots  (N_{l_n}+1)\}]\nonumber\\
&<& 2 E(l_n \ln (n/l_n))\nonumber\\
&\propto & \frac{n \ln \ln n}{\ln n}\quad {\rm as}~n\rightarrow \infty,
\end{eqnarray}
where $N_1+1+\cdots +N_{l_n}+1=n$.
This means subexponential instability, $E(X_1+\cdots +X_n)=o(n)$.

%


\begin{thebibliography}{10}%
\makeatletter
\providecommand \@ifxundefined [1]{%
 \ifx #1\undefined \expandafter \@firstoftwo
 \else \expandafter \@secondoftwo
\fi
}%
\providecommand \@ifnum [1]{%
 \ifnum #1\expandafter \@firstoftwo
 \else \expandafter \@secondoftwo
\fi
}%
\providecommand \enquote [1]{``#1''}%
\providecommand \bibnamefont  [1]{#1}%
\providecommand \bibfnamefont [1]{#1}%
\providecommand \citenamefont [1]{#1}%
\providecommand\href[0]{\@sanitize\@href}%
\providecommand\@href[1]{\endgroup\@@startlink{#1}\endgroup\@@href}%
\providecommand\@@href[1]{#1\@@endlink}%
\providecommand \@sanitize [0]{\begingroup\catcode`\&12\catcode`\#12\relax}%
\@ifxundefined \pdfoutput {\@firstoftwo}{%
 \@ifnum{\z@=\pdfoutput}{\@firstoftwo}{\@secondoftwo}%
}{%
 \providecommand\@@startlink[1]{\leavevmode}%
 \providecommand\@@endlink[0]{}%
}{%
 \providecommand\@@startlink[1]{%
  \leavevmode
  \pdfstartlink
   attr{/Border[0 0 1 ]/H/I/C[0 1 1]}%
   user{/Subtype/Link/A<</Type/Action/S/URI/URI(#1)>>}%
  \relax
 }%
 \providecommand\@@endlink[0]{\pdfendlink}%
}%
\providecommand \url  [0]{\begingroup\@sanitize \@url }%
\providecommand \@url [1]{\endgroup\@href {#1}{\urlprefix}}%
\providecommand \urlprefix [0]{URL }%
\providecommand \Eprint[0]{\href }%
\@ifxundefined \urlstyle {%
  \providecommand \doi [1]{doi:\discretionary{}{}{}#1}%
}{%
  \providecommand \doi [0]{doi:\discretionary{}{}{}\begingroup
  \urlstyle{rm}\Url }%
}%
\providecommand \doibase [0]{http://dx.doi.org/}%
\providecommand \Doi[1]{\href{\doibase#1}}%
\providecommand \selectlanguage [0]{\@gobble}%
\providecommand \bibinfo [0]{\@secondoftwo}%
\providecommand \bibfield [0]{\@secondoftwo}%
\providecommand \translation [1]{[#1]}%
\providecommand \BibitemOpen[0]{}%
\providecommand \bibitemStop [0]{}%
\providecommand \bibitemNoStop [0]{.\EOS\space}%
\providecommand \EOS [0]{\spacefactor3000\relax}%
\providecommand \BibitemShut [1]{\csname bibitem#1\endcsname}%
\bibitem{Dorfmann1999}%
  \BibitemOpen
  \bibfield{author}{%
  \bibinfo {author} {\bibfnamefont{J.~R.}\ \bibnamefont{Dorfmann}},\ }%
  \emph{\bibinfo {title} {An Introduction to Chaos in Nonequilibrium
  Statistical Mechanics}}\ (\bibinfo {publisher} {Cambridge Unviersity Press},\
  \bibinfo {year} {1999})\BibitemShut{NoStop}%
\bibitem{Golding2006}%
  \BibitemOpen
  \bibfield{author}{%
  \bibinfo {author} {\bibfnamefont{I.}~\bibnamefont{Golding}}\ and\ \bibinfo
  {author} {\bibfnamefont{E.~C.}\ \bibnamefont{Cox}},\ }%
  \bibfield{title}{%
  \enquote{\bibinfo {title} {Physcial nature of bacterial cytoplasm},}\ }%
  \bibfield{journal}{%
  \bibinfo {journal} {Phys. Rev. Lett.}\ }%
  \textbf{\bibinfo {volume} {96}},\ \bibinfo {pages} {098102} (\bibinfo {year}
  {2006})\BibitemShut{NoStop}%
\bibitem{Brok2003}%
  \BibitemOpen
  \bibfield{author}{%
  \bibinfo {author} {\bibfnamefont{X.}~\bibnamefont{Brokmann}}, \bibinfo
  {author} {\bibfnamefont{J.-P.}\ \bibnamefont{Hermier}}, \bibinfo {author}
  {\bibfnamefont{G.}~\bibnamefont{Messin}}, \bibinfo {author}
  {\bibfnamefont{P.}~\bibnamefont{Desbiolles}}, \bibinfo {author}
  {\bibfnamefont{J.-P.}\ \bibnamefont{Bouchaud}},\ and\ \bibinfo {author}
  {\bibfnamefont{M.}~\bibnamefont{Dahan}},\ }%
  \bibfield{title}{%
  \enquote{\bibinfo {title} {Statistical aging and nonergodicity in the
  fluorescence of single nanocrystals},}\ }%
  \bibfield{journal}{%
  \bibinfo {journal} {Phys. Rev. Lett.}\ }%
  \textbf{\bibinfo {volume} {90}},\ \bibinfo {pages} {120601} (\bibinfo {year}
  {2003})\BibitemShut{NoStop}%
\bibitem{He2008}%
  \BibitemOpen
  \bibfield{author}{%
  \bibinfo {author} {\bibfnamefont{Y.}~\bibnamefont{He}}, \bibinfo {author}
  {\bibfnamefont{S.}~\bibnamefont{Burov}}, \bibinfo {author}
  {\bibfnamefont{R.}~\bibnamefont{Metzler}},\ and\ \bibinfo {author}
  {\bibfnamefont{E.}~\bibnamefont{Barkai}},\ }%
  \bibfield{title}{%
  \enquote{\bibinfo {title} {Random time-scale invariant diffusion and
  transport coefficients},}\ }%
  \bibfield{journal}{%
  \bibinfo {journal} {Phys. Rev. Lett.}\ }%
  \textbf{\bibinfo {volume} {101}},\ \bibinfo {pages} {058101} (\bibinfo {year}
  {2008})\BibitemShut{NoStop}%
\bibitem{Akimoto2008}%
  \BibitemOpen
  \bibfield{author}{%
  \bibinfo {author} {\bibfnamefont{T.}~\bibnamefont{Akimoto}},\ }%
  \bibfield{title}{%
  \enquote{\bibinfo {title} {Generalized arcsine law and stable law in an
  infinite measure dynamical system},}\ }%
  \bibfield{journal}{%
  \bibinfo {journal} {J. Stat. Phys.}\ }%
  \textbf{\bibinfo {volume} {132}},\ \bibinfo {pages} {171--186} (\bibinfo
  {year} {2008})\BibitemShut{NoStop}%
\bibitem{Geisel1984}%
  \BibitemOpen
  \bibfield{author}{%
  \bibinfo {author} {\bibfnamefont{T.}~\bibnamefont{Geisel}}\ and\ \bibinfo
  {author} {\bibfnamefont{S.}~\bibnamefont{Thomae}},\ }%
  \bibfield{title}{%
  \enquote{\bibinfo {title} {Anomalous diffusion in intermittent chaotic
  systems},}\ }%
  \bibfield{journal}{%
  \bibinfo {journal} {Phys. Rev. Lett.}\ }%
  \textbf{\bibinfo {volume} {52}},\ \bibinfo {pages} {1936--1939} (\bibinfo
  {year} {1984})\BibitemShut{NoStop}%
\bibitem{Barkai2003}%
  \BibitemOpen
  \bibfield{author}{%
  \bibinfo {author} {\bibfnamefont{E.}~\bibnamefont{Barkai}},\ }%
  \bibfield{title}{%
  \enquote{\bibinfo {title} {Aging in subdiffusion generated by a deterministic
  dynamical system},}\ }%
  \bibfield{journal}{%
  \bibinfo {journal} {Phys. Rev. Lett.}\ }%
  \textbf{\bibinfo {volume} {90}},\ \bibinfo {pages} {104101} (\bibinfo {year}
  {2003})\BibitemShut{NoStop}%
\bibitem{Aizawa1989a}%
  \BibitemOpen
  \bibfield{author}{%
  \bibinfo {author} {\bibfnamefont{Y.}~\bibnamefont{Aizawa}}, \bibinfo {author}
  {\bibfnamefont{Y.}~\bibnamefont{Kikuchi}}, \bibinfo {author}
  {\bibfnamefont{T.}~\bibnamefont{Harayama}}, \bibinfo {author}
  {\bibfnamefont{K.}~\bibnamefont{Yamamoto}}, \bibinfo {author}
  {\bibfnamefont{M.}~\bibnamefont{Ota}},\ and\ \bibinfo {author}
  {\bibfnamefont{K.}~\bibnamefont{Tanaka}},\ }%
  \bibfield{title}{%
  \enquote{\bibinfo {title} {Stagnant motions in hamiltonian systems},}\ }%
  \bibfield{journal}{%
  \bibinfo {journal} {Prog. Theor. Phys. Suppl.}\ }%
  \textbf{\bibinfo {volume} {98}},\ \bibinfo {pages} {36--82} (\bibinfo {year}
  {1989})\BibitemShut{NoStop}%
\bibitem{Akimoto2008a}%
  \BibitemOpen
  \bibfield{author}{%
  \bibinfo {author} {\bibfnamefont{T.}~\bibnamefont{Akimoto}},\ }%
  \bibfield{title}{%
  \enquote{\bibinfo {title} {On the definition of equilibrium and
  non-equilibrium states in dynamical systems},}\ }%
  \bibfield{journal}{%
  \bibinfo {journal} {AIP Conf. Proc.}\ }%
  \textbf{\bibinfo {volume} {1076}},\ \bibinfo {pages} {5--8} (\bibinfo {year}
  {2008})\BibitemShut{NoStop}%
\bibitem{Korabel2009}%
  \BibitemOpen
  \bibfield{author}{%
  \bibinfo {author} {\bibfnamefont{N.}~\bibnamefont{Korabel}}\ and\ \bibinfo
  {author} {\bibfnamefont{E.}~\bibnamefont{Barkai}},\ }%
  \bibfield{title}{%
  \enquote{\bibinfo {title} {Pesin-type identity for intermittent dynamics with
  a zero lyapunov exponent},}\ }%
  \bibfield{journal}{%
  \bibinfo {journal} {Phys. Rev. Lett.}\ }%
  \textbf{\bibinfo {volume} {102}},\ \bibinfo {pages} {050601} (\bibinfo {year}
  {2009})\BibitemShut{NoStop}%
\bibitem{Aaronson1981}%
  \BibitemOpen
  \bibfield{author}{%
  \bibinfo {author} {\bibfnamefont{J.}~\bibnamefont{Aaronson}},\ }%
  \bibfield{title}{%
  \enquote{\bibinfo {title} {The asymptotic distributional behavior of
  transformations preserving infinite measures},}\ }%
  \bibfield{journal}{%
  \bibinfo {journal} {J. D'Analyse Math.}\ }%
  \textbf{\bibinfo {volume} {39}},\ \bibinfo {pages} {203--234} (\bibinfo
  {year} {1981})\BibitemShut{NoStop}%
\bibitem{Birkhoff1931}%
  \BibitemOpen
  \bibfield{author}{%
  \bibinfo {author} {\bibfnamefont{G.~D.}\ \bibnamefont{Birkhoff}},\ }%
  \bibfield{title}{%
  \enquote{\bibinfo {title} {Proof of the ergodic theorem},}\ }%
  \bibfield{journal}{%
  \bibinfo {journal} {Proc. Natl. Acad. Sci. USA}\ }%
  \textbf{\bibinfo {volume} {17}},\ \bibinfo {pages} {656--660} (\bibinfo
  {year} {1931})\BibitemShut{NoStop}%
\bibitem{Gaspard1988}%
  \BibitemOpen
  \bibfield{author}{%
  \bibinfo {author} {\bibfnamefont{P.}~\bibnamefont{Gaspard}}\ and\ \bibinfo
  {author} {\bibfnamefont{X.~J.}\ \bibnamefont{Wang}},\ }%
  \bibfield{title}{%
  \enquote{\bibinfo {title} {Sporadicity: Between periodic and chaotic
  dynamical behaviors},}\ }%
  \bibfield{journal}{%
  \bibinfo {journal} {Proc. Natl. Acad. Sci. USA}\ }%
  \textbf{\bibinfo {volume} {85}},\ \bibinfo {pages} {4591--4595} (\bibinfo
  {year} {1988})\BibitemShut{NoStop}%
\bibitem{Darling1957}%
  \BibitemOpen
  \bibfield{author}{%
  \bibinfo {author} {\bibfnamefont{D.~A.}\ \bibnamefont{Darling}}\ and\
  \bibinfo {author} {\bibfnamefont{M.}~\bibnamefont{Kac}},\ }%
  \bibfield{title}{%
  \enquote{\bibinfo {title} {On occupation times for markov processes},}\ }%
  \bibfield{journal}{%
  \bibinfo {journal} {Trans. Am. Math. Soc.}\ }%
  \textbf{\bibinfo {volume} {84}},\ \bibinfo {pages} {444--458} (\bibinfo
  {year} {1957})\BibitemShut{NoStop}%
\bibitem{FN1}%
  \BibitemOpen
  \bibinfo {note} {$\int_X |f| dm<\infty$.}\BibitemShut{Stop}%
\bibitem{FN2}%
  \BibitemOpen
  \bibinfo {note} {The random variable $Y_{\alpha}$ on $\R$ has the normalized
  Mittag-Leffler distribution of order $\alpha$ if \[
  E(e^{zY_{\alpha}})=\sum_{k=0}^{\infty}\frac{\Gamma(1+\alpha)^kz^k}{\Gamma(1+%
k\alpha)}, \] where $E(\cdot )$ is the expectation.}\BibitemShut{Stop}%
\bibitem{Aaronson1997}%
  \BibitemOpen
  \bibfield{author}{%
  \bibinfo {author} {\bibfnamefont{J.}~\bibnamefont{Aaronson}},\ }%
  \emph{\bibinfo {title} {An Introduction to Infinite Ergodic Theory}}\
  (\bibinfo {publisher} {American Mathematical Society},\ \bibinfo {address}
  {Province},\ \bibinfo {year} {1997})\BibitemShut{NoStop}%
\bibitem{Milnor1985}%
  \BibitemOpen
  \bibfield{author}{%
  \bibinfo {author} {\bibfnamefont{J.}~\bibnamefont{Milnor}},\ }%
  \bibfield{title}{%
  \enquote{\bibinfo {title} {On the concept of attractor},}\ }%
  \bibfield{journal}{%
  \bibinfo {journal} {Commun. Math. Phys.}\ }%
  \textbf{\bibinfo {volume} {99}},\ \bibinfo {pages} {177--195} (\bibinfo
  {year} {1985})\BibitemShut{NoStop}%
\bibitem{Grassberger1981}%
  \BibitemOpen
  \bibfield{author}{%
  \bibinfo {author} {\bibfnamefont{P.}~\bibnamefont{Grassberger}}\ and\
  \bibinfo {author} {\bibfnamefont{M.}~\bibnamefont{Scheunert}},\ }%
  \bibfield{title}{%
  \enquote{\bibinfo {title} {Some more universal scaling laws for critical
  mappings},}\ }%
  \bibfield{journal}{%
  \bibinfo {journal} {J. Stat. Phys.}\ }%
  \textbf{\bibinfo {volume} {26}},\ \bibinfo {pages} {697--717} (\bibinfo
  {year} {1981})\BibitemShut{NoStop}%
\bibitem{Collet1980}%
  \BibitemOpen
  \bibfield{author}{%
  \bibinfo {author} {\bibfnamefont{P.}~\bibnamefont{Collet}}\ and\ \bibinfo
  {author} {\bibfnamefont{J.~P.}\ \bibnamefont{Eckmann}},\ }%
  \emph{\bibinfo {title} {Iterated Maps on the Interval as Dynamical Systems}}\
  (\bibinfo {publisher} {Birk\"{a}user, Boston},\ \bibinfo {year}
  {1980})\BibitemShut{NoStop}%
\bibitem{FN3}%
  \BibitemOpen
  \bibinfo {note} {Here we assume that there exists the only one indifferent
  fixed point. However, one can show in a similar way when there exist some
  indifferent fixed points or indifferent periodic points.}\BibitemShut{Stop}%
  \bibitem{Thaler1983}%
  \BibitemOpen
  \bibfield{author}{%
  \bibinfo {author} {\bibfnamefont{M.}\ \bibnamefont{Thaler}},\ }%
  \bibfield{title}{%
  \enquote{\bibinfo {title} {Transformations on [0,1] with infinite invariant
  measures},}\ }%
  \bibfield{journal}{%
  \bibinfo {journal} {Isr. J. Math.}\ }%
  \textbf{\bibinfo {volume} {46}},\ \bibinfo {pages} {67--96} (\bibinfo {year}
  {1983})\BibitemShut{NoStop}%
\bibitem{FN4}%
  \BibitemOpen
  \bibinfo {note} {In the case of $|X_1|+\cdots +|X_n| =o(n)$, one can show in
  the same way that the invariant measure is an infinite
  one.}\BibitemShut{Stop}%
\bibitem{Feller1971}%
  \BibitemOpen
  \bibfield{author}{%
  \bibinfo {author} {\bibfnamefont{W.}~\bibnamefont{Feller}},\ }%
  \emph{\bibinfo {title} {An Introduction to Probability Theory and its
  Applications}},\ \bibinfo {edition} {2nd}\ ed.,\ Vol.~\bibinfo {volume} {2}\
  (\bibinfo {publisher} {Wiley, New York},\ \bibinfo {year}
  {1971})\BibitemShut{NoStop}%
\bibitem{Inoue1992}%
  \BibitemOpen
  \bibfield{author}{%
  \bibinfo {author} {\bibfnamefont{T.}~\bibnamefont{Inoue}},\ }%
  \bibfield{title}{%
  \enquote{\bibinfo {title} {Weakly attracting repellors for piecewise convex
  maps},}\ }%
  \bibfield{journal}{%
  \bibinfo {journal} {Japan J. Indust. Appl. Math.}\ }%
  \textbf{\bibinfo {volume} {9}},\ \bibinfo {pages} {413--430} (\bibinfo {year}
  {1992})\BibitemShut{NoStop}%
\bibitem{Zweimuller2004}%
  \BibitemOpen
  \bibfield{author}{%
  \bibinfo {author} {\bibfnamefont{R.}~\bibnamefont{Zweim{\"u}ller}},\ }%
  \bibfield{title}{%
  \enquote{\bibinfo {title} {S-unimodal misiurewicz maps with flat critical
  points},}\ }%
  \bibfield{journal}{%
  \bibinfo {journal} {Fund. Math.}\ }%
  \textbf{\bibinfo {volume} {181}},\ \bibinfo {pages} {1} (\bibinfo {year}
  {2004})\BibitemShut{NoStop}%
\bibitem{Galatolo2003}%
  \BibitemOpen
  \bibfield{author}{%
  \bibinfo {author} {\bibfnamefont{S.}~\bibnamefont{Galatolo}},\ }%
  \bibfield{title}{%
  \enquote{\bibinfo {title} {Complexity, initial condition sensitivity,
  dimension and weak chaos in dynamical systems},}\ }%
  \bibfield{journal}{%
  \bibinfo {journal} {Nonlinearity}\ }%
  \textbf{\bibinfo {volume} {16}},\ \bibinfo {pages} {1219--1238} (\bibinfo
  {year} {2003})\BibitemShut{NoStop}%
\bibitem{Bonanno2004}%
  \BibitemOpen
  \bibfield{author}{%
  \bibinfo {author} {\bibfnamefont{C.}~\bibnamefont{Bonanno}}\ and\ \bibinfo
  {author} {\bibfnamefont{S.}~\bibnamefont{Galatolo}},\ }%
  \bibfield{title}{%
  \enquote{\bibinfo {title} {Algorithmic information for interval maps with an
  indifferent fixed point and infinite invariant measure},}\ }%
  \bibfield{journal}{%
  \bibinfo {journal} {Chaos}\ }%
  \textbf{\bibinfo {volume} {14}},\ \bibinfo {pages} {756--762} (\bibinfo
  {year} {2004})\BibitemShut{NoStop}%
\bibitem{Wang2004}%
  \BibitemOpen
  \bibfield{author}{%
  \bibinfo {author} {\bibfnamefont{X.}~\bibnamefont{Wang}}, \bibinfo {author}
  {\bibfnamefont{M.}~\bibnamefont{Zhan}}, \bibinfo {author}
  {\bibfnamefont{C.}~\bibnamefont{Lai}},\ and\ \bibinfo {author}
  {\bibfnamefont{Y.}~\bibnamefont{Lai}},\ }%
  \bibfield{title}{%
  \enquote{\bibinfo {title} {Strange nonchaotic attractors in random dynamical
  systems},}\ }%
  \bibfield{journal}{%
  \bibinfo {journal} {Phys. Rev. Lett.}\ }%
  \textbf{\bibinfo {volume} {92}},\ \bibinfo {pages} {074102} (\bibinfo {year}
  {2004})\BibitemShut{NoStop}%
\bibitem{Ashwin1998}%
  \BibitemOpen
  \bibfield{author}{%
  \bibinfo {author} {\bibfnamefont{P.}~\bibnamefont{Ashwin}}, \bibinfo {author}
  {\bibfnamefont{P.~J.}\ \bibnamefont{Aston}},\ and\ \bibinfo {author}
  {\bibfnamefont{M.}~\bibnamefont{Nicol}},\ }%
  \bibfield{title}{%
  \enquote{\bibinfo {title} {On the unfolding of a blowout bifurcation},}\ }%
  \bibfield{journal}{%
  \bibinfo {journal} {Physica D}\ }%
  \textbf{\bibinfo {volume} {111}},\ \bibinfo {pages} {81--95} (\bibinfo {year}
  {1998})\BibitemShut{NoStop}%
\bibitem{Badii1988}%
  \BibitemOpen
  \bibfield{author}{%
  \bibinfo {author} {\bibfnamefont{R.}~\bibnamefont{Badii}}, \bibinfo {author}
  {\bibfnamefont{K.}~\bibnamefont{Heinzelmann}}, \bibinfo {author}
  {\bibfnamefont{P.~F.}\ \bibnamefont{Meier}},\ and\ \bibinfo {author}
  {\bibfnamefont{A.}~\bibnamefont{Politi}},\ }%
  \bibfield{title}{%
  \enquote{\bibinfo {title} {Correlation functions and generalized lyapunov
  exponents},}\ }%
  \bibfield{journal}{%
  \bibinfo {journal} {Phys. Rev. A}\ }%
  \textbf{\bibinfo {volume} {37}},\ \bibinfo {pages} {1323--1328} (\bibinfo
  {year} {1988})\BibitemShut{NoStop}%
\bibitem{Fujisaka1983}%
  \BibitemOpen
  \bibfield{author}{%
  \bibinfo {author} {\bibfnamefont{H.}~\bibnamefont{Fujisaka}},\ }%
  \bibfield{title}{%
  \enquote{\bibinfo {title} {Statistical dynamics generated by fluctuations of
  local lyapunov exponents},}\ }%
  \bibfield{journal}{%
  \bibinfo {journal} {Prog. Theor. Phys.}\ }%
  \textbf{\bibinfo {volume} {70}},\ \bibinfo {pages} {1264--1275} (\bibinfo
  {year} {1983})\BibitemShut{NoStop}%
\bibitem{Mori1989}%
  \BibitemOpen
  \bibfield{author}{%
  \bibinfo {author} {\bibfnamefont{H.}~\bibnamefont{Mori}}, \bibinfo {author}
  {\bibfnamefont{H.}~\bibnamefont{Hata}}, \bibinfo {author}
  {\bibfnamefont{T.}~\bibnamefont{Horita}},\ and\ \bibinfo {author}
  {\bibfnamefont{T.}~\bibnamefont{Kobayashi}},\ }%
  \bibfield{title}{%
  \enquote{\bibinfo {title} {Statistical mechanics of dynamical systems},}\ }%
  \bibfield{journal}{%
  \bibinfo {journal} {Prog. Theor. Phys. Suppl.}\ }%
  \textbf{\bibinfo {volume} {99}},\ \bibinfo {pages} {1--63} (\bibinfo {year}
  {1989})\BibitemShut{NoStop}%
\bibitem{Kuzovkov2002}%
  \BibitemOpen
  \bibfield{author}{%
  \bibinfo {author} {\bibfnamefont{V.~N.}\ \bibnamefont{Kuzovkov}}, \bibinfo
  {author} {\bibfnamefont{W.}~\bibnamefont{von Niessen}}, \bibinfo {author}
  {\bibfnamefont{V.}~\bibnamefont{Kashcheyevs}},\ and\ \bibinfo {author}
  {\bibfnamefont{O.}~\bibnamefont{Hein}},\ }%
  \bibfield{title}{%
  \enquote{\bibinfo {title} {Exact analytic solution for the generalized
  lyapunov exponent of the two-dimensional anderson localization},}\ }%
  \bibfield{journal}{%
  \bibinfo {journal} {J. Phys.: Condens. Matter}\ }%
  \textbf{\bibinfo {volume} {14}},\ \bibinfo {pages} {13777--13797} (\bibinfo
  {year} {2002})\BibitemShut{NoStop}%
\end{thebibliography}

\end{document}